\documentclass[longauth]{aa}
\usepackage{amsmath}
\usepackage{amssymb}
\usepackage{blindtext}
\usepackage{bm}
\usepackage{color}
\usepackage{graphicx}
\usepackage{hyperref}
\usepackage{lipsum}
\usepackage{longtable}
\usepackage{multicol}
\usepackage{multirow}
\usepackage{natbib}
\usepackage{txfonts}
\usepackage{wasysym}
\usepackage[capitalize,nameinlink]{cleveref}
\usepackage{threeparttable}
\usepackage{booktabs}

\creflabelformat{equation}{#2#1#3}
\crefrangelabelformat{equation}{#3#1#4 to #5#2#6}

\hypersetup{colorlinks,
  linkcolor=red,
  filecolor=red,
  urlcolor=magenta,
  citecolor=blue}

\usepackage{txfonts}

\usepackage{xspace}

\usepackage[dvipsnames]{xcolor}

\usepackage{enumitem}
\usepackage{orcidlink}

\begin{document}

   \title{The role of supercluster filaments in shaping galaxy clusters}
   \titlerunning{Filaments shaping clusters}
        \author{Ra\'ul Baier-Soto\orcidlink{0009-0008-4255-1309}\inst{1,2,3}\thanks{\email{rbaier@usm.cl}}
        \and
        Yara Jaffé\orcidlink{0000-0003-2150-1130} \inst{1,3}
        \and
       Alexis Finoguenov\orcidlink{0000-0002-4606-5403}\inst{4}
       \and
        Christopher P. Haines\orcidlink{0000-0002-8814-8960}\inst{5,3}
        \and  Paola Merluzzi\orcidlink{0000-0003-3966-2397}\inst{6}
        \and Hugo Méndez-Hernández\orcidlink{0000-0003-3057-9677}\inst{7,3}
        \and Antonela Monachesi\orcidlink{0000-0003-2325-9616}\inst{7}
        \and Ulrike Kuchner\orcidlink{0000-0002-0035-5202}\inst{8}
        \and Rory Smith\orcidlink{0000-0001-5303-6830}\inst{1,3}
        \and Nicolas Tejos\orcidlink{0000-0002-1883-4252}\inst{2}
        \and Cristóbal Sifón\orcidlink{0000-0002-8149-1352}\inst{2}
        \and Maria Argudo-Fernández\orcidlink{0000-0002-0789-2326}\inst{9,10}
        \and C. R. Bom\orcidlink{0000-0003-4383-2969}\inst{11}
        \and Johan Comparat\orcidlink{0000-0001-9200-1497}\inst{12}
        \and Ricardo Demarco\orcidlink{0000-0003-3921-2177}\inst{13}
        \and Rodrigo F. Haack\orcidlink{0009-0005-6830-1832}\inst{14,15}
        \and Ivan Lacerna\orcidlink{0000-0002-7802-7356}\inst{5}
        \and  E. V. R. Lima\orcidlink{0000-0002-6268-8600}\inst{15}
        \and Ciria Lima-Dias\orcidlink{0009-0006-0373-8168}\inst{7}
        \and Elismar Lösch\orcidlink{0000-0003-2561-0756}\inst{16}
        \and C. Mendes de Oliveira\orcidlink{0000-0002-5267-9065}\inst{16}
        \and Diego Pallero\orcidlink{0000-0002-1577-7475}\inst{1,3}
        \and Laerte Sodré Jr\orcidlink{0000-0002-3876-268X}\inst{16}
        \and Gabriel S. M. Teixeira\orcidlink{0000-0002-1594-208X}\inst{11}
        \and O. Alghamdi\inst{17}
        \and F. Almeida-Fernandes \orcidlink{0000-0002-8048-8717}\inst{16,18}
        \and Stefania Barsanti\orcidlink{0000-0002-9332-5386}\inst{19}
        \and Lawrence E. Bilton\inst{20}
        \and M. Canducci\orcidlink{0000-0003-2264-9743}\inst{17}
        \and Maiara Carvalho\orcidlink{0009-0008-9042-4478}\inst{16}
        \and Giuseppe D’Ago\orcidlink{0000-0001-9697-7331}\inst{21}
        \and Alexander Fritz\inst{22}
        \and Fábio R. Herpich\orcidlink{0000-0001-7907-7884}\inst{23}
        \and E. Ibar\orcidlink{0009-0008-9801-2224}\inst{24,3}
        \and Hyowon Kim\orcidlink{0000-0003-4032-8572}\inst{1,3}
        \and  Sebastian Lopez\orcidlink{0000-0003-0389-0902}\inst{25}
        \and Alessia Moretti\orcidlink{0000-0002-1688-482X}\inst{26}
        \and L. M. I. Nakazono\orcidlink{0000-0001-6480-1155}\inst{16}
        \and D. E. Olave-Rojas\inst{27}
        \and G. B. Oliveira Schwarz\inst{28}
        \and Franco Piraino-Cerda\orcidlink{0009-0008-0197-3337}\inst{1,2,3}
        \and Emanuela Pompei\orcidlink{0000-0001-7578-8160}\inst{29}
        \and U. Rescigno\orcidlink{0000-0002-9280-0173}\inst{5}
        \and Boudewijn F. Roukema\orcidlink{0000-0002-3772-0250} \inst{30,31}
        \and V. M. Sampaio\orcidlink{0000-0001-6556-637X}\inst{1,3}
        \and P. Tiño\orcidlink{0000-0003-2330-128X}\inst{17}
        \and P. Vásquez-Bustos\orcidlink{0000-0001-7140-3980}\inst{9}
  }

   \institute{
        Departamento de Física, Universidad Técnica Federico Santa María, Avenida España 1680, Valparaíso, Chile
        \and Instituto de F\'isica, Pontificia Universidad Católica de Valparaíso, Casilla 4059, Valparaíso, Chile
        \and Millennium Nucleus for Galaxies (MINGAL)
        \and Department of Physics, University of Helsinki, Gustaf Hällströmin katu 2, 00560 Helsinki, Finland
        \and Instituto de Astronom\'ia y Ciencias Planetarias (INCT), Universidad de Atacama, Copayapu 485, Copiap\'o, Chile\
        \and INAF - Osservatorio Astronomico di Capodimonte, Salita
        Moiariello 16 80131, Napoli, Italy
        \and Departamento de Astronomía, Universidad de La Serena, Avda.
        Raúl Bitrán 1305, La Serena, Chile
        \and School of Physics \& Astronomy, University of Nottingham, Nottingham NG7 2RD, UK     
        \and Departamento de Física Teórica y del Cosmos, Edificio Mecenas,
        Campus Fuentenueva, Universidad de Granada, E-18071, Granada,
        Spain
        \and  Instituto Universitario Carlos I de Física Teórica y Computacional, Universidad de Granada, 18071 Granada, Spain
        \and Centro Brasileiro de Pesquisas Físicas, Rua Dr. Xavier Sigaud 150, 22290-180 Rio de Janeiro, RJ, Brazil
        \and{Max-Planck-Institut für extraterrestrische Physik (MPE), Gießenbachstraße 1, 85748 Garching bei München, Germany}
        \and Institute of Astrophysics, Facultad de Ciencias Exactas, Universidad
        Andrés Bello, Sede Concepción, Talcahuano, Chile
        \and Instituto de Astrofísica de La Plata, CONICET-UNLP, Paseo del Bosque s/n, B1900FWA, Argentina
       \and Facultad de Ciencias Astronómicas y Geofísicas, Universidad Nacional de La Plata, Paseo del Bosque s/n, B1900FWA, Argentina
        \and  Departamento de Astronomia, Instituto de Astronomia, Geofísica e Ciências Atmosféricas, Universidade de São Paulo, Rua do Matão 1226, Cidade Universitária, São Paulo 05508-090, Brazil
        \and School of Computer Science, University of Birmingham, Edgbaston, Birmingham B15 2TT, UK
        \and Observatório do Valongo, Ladeira Pedro Antônio, 43, Saúde, Rio de Janeiro 20080-090, BR, Brazil
        \and Sydney Institute for Astronomy, School of Physics, University of Sydney, NSW 2006, Australia
        \and Centre of Excellence for Data Science, Artificial Intelligence \&
        Modelling, The University of Hull, Cottingham Road, KingstonUpon-Hull, HU6 7RX, UK
        \and Institute of Astronomy, University of Cambridge, Madingley Road, Cambridge CB3 0HA, United Kingdom
        \and Kuffner Observatory, Johann-Staud-Straße 10, 1160 Vienna, Austria
        \and Laboratório Nacional de Astrofísica (LNA/MCTI), Rua Estados Unidos, 154, Itajubá 37504-364, Brazil
        \and Instituto de F\'isica y Astronom\'ia, Universidad de Valpara\'iso, Avda. Gran Breta\~na 1111, Valpara\'iso, Chile
        \and Departamento de Astronomía, Universidad de Chile, Casilla 36-D, Santiago, Chile
        \and INAF-Padova Astronomical Observatory, Vicolo dell’Osservatorio
        5, I-35122 Padova, Italy
        \and Departamento de Tecnologías Industriales, Facultad de Ingeniería, Universidad de Talca, Los Niches km 1, Curicó, Chile
        \and Universidade Presbiteriana Mackenzie, R. da Consola\c{c}ao, 930 -Consola\c{c}ao, S\~ao Paulo, Brazil
        \and European Southern Observatory, Science Operations, Alonso de Cordova 3107, Vitacura, 19001 Santiago, Chile
        \and Institute of Astronomy, Faculty of Physics,
       Astronomy and Informatics, Nicolaus Copernicus
       University, Grudziadzka 5, 87-100 Toru\'n, Poland
        \and  Univ Lyon, Ens de Lyon, Univ Lyon1, CNRS, Centre de
           Recherche Astrophysique de Lyon UMR5574, F--69007, Lyon,
           France
 }

   \date{Received ; accepted }

  \abstract
   {In a hierarchical $\Lambda$CDM Universe, cosmic filaments serve as the primary channels for matter accretion into galaxy clusters, influencing the shape of their dark matter halos.}
   {We investigate whether the elongation of galaxy clusters correlates with the orientation of surrounding filaments, providing the first observational test of this relationship in large supercluster regions.}
   {We identified and characterized cosmic filaments in two dimensions within the two superclusters that are part of the low-redshift sub-survey of the Chilean Cluster Galaxy Evolution Survey (CHANCES): the Shapley supercluster and the Horologium-Reticulum supercluster. We analyzed the alignment between filament directions—traced by galaxy distributions—and the triaxiality of cluster gravitational potentials—traced by X-ray emission—using publicly available optical and X-ray data.}
   {We have found that most (82\%) of the X-ray clusters are associated with and interconnected by the optically detected filaments. The clusters-filaments alignment analysis shows that the elongation of most clusters is well aligned with nearby filaments, providing observational confirmation of theoretical predictions, with the alignment progressively reducing at larger cluster-centric distances ($>$ 1.6 $r_{200}$).} 
   {Overall, our results support the notion that filaments are the main source of galaxy accretion at {redshift below 0.1}, and additionally provide evidence that matter accretion through filaments shapes the gravitational potential of galaxy clusters. We propose this measurement as a simple observational proxy to determine the direction of accretion in clusters, which is key to understanding both galaxy evolution and the merger history of galaxy clusters.}


   \maketitle

\section{Introduction}

The large-scale structure (LSS) of the Universe is composed by an intricate and inhomogeneous distribution of matter known as the cosmic web. This structure is constituted by four main components: nodes (also referred to as galaxy clusters) containing up to thousands of galaxies in a sphere of $\sim 1 - 2$ Mpc radius, filaments, sheet-like walls, and large-scale voids spanning up to hundreds of {megaparsecs} in size \citep[e.g.,][]{Cautun2014}. This web-like structure {has been shaped by} the anisotropic nature of the gravitational collapse \citep{Peebles1980}, following a well-defined sequence that {began with regions collapsing along one axis} to form walls, then {along another} axis to form filaments, and finally {collapsing} in all directions to form nodes \citep[e.g.,][]{zel1970gravitational,arnold1982large,shandarin1984rich}.  

Although filaments occupy less than ten percent of the total volume of the present Universe, they contain more than half of the {mass} \citep[e.g.,][]{Cautun2014, Tempel, cui2019large, martizzi2019baryons}, {and are the main channels through which it is transported across the cosmic web.} The intersections of filaments host galaxy clusters, where the mass density $1 + \delta$\footnote{For simplicity, \cite{Cautun2014} computed the density $\rho$ in units of the mean background density $\overline{\rho}$ as $1+\delta = \rho/\overline{\rho}$.} exceeds $\sim 100$ the background average density \citep[e.g.,][]{Cautun2014}. Clusters are thought to grow from the accretion of dark matter, gas, and galaxies. If this accretion happens mainly through cosmic filaments (as suggested, for example, by \cite{kravtsov2012formation}, \cite{Cautun2014} and \cite{umehata2019gas}), a connection is expected between the cluster geometry, its gravitational potential, and the orientation of its associated filaments. 

Indeed, the flow of matter into galaxy clusters is highly anisotropic \citep[e.g.,][]{gouin2021shape}, rather than
uniformly from all sides, indicating preferential directions along which accretion occurs through the filaments, and several studies have focused on studying this correlation on isolated systems using different indicators. Considering that cluster members tend to be distributed preferentially in line with the position angle (PA) of the major axis of the brightest cluster galaxy (BCG) \citep[e.g.,][]{sastry1968clusters}, \cite{smith2023bcg} studied the relation between the position angle of the BCG and the projected orientation of the filaments connected to the clusters, founding a strong sign of alignment between the distribution of galaxies and the PA of the BCG, as well as with the LSS filaments around the clusters, even at distances of $\sim$ 10 Mpc. From the theoretical side, using 324 simulations of massive clusters and their surrounding environment from \textsc{The ThreeHundred} project \citep{cui2018three,mostoghiu2019three,wang2018three,arthur2019thethreehundred,ansarifard2020three,santoni2024three}, \cite{Kuchner2020} have also shown strong signs of alignment between the major axis of the cluster dark-matter halos and the connected filaments.

Cosmic filaments play a key role in shaping the structure of galaxy clusters through anisotropic and directional matter accretion. As a result, the asphericity of clusters is a natural consequence of their alignment with these filaments, and reflects their relatively recent formation and the less evolved and relaxed state of their dark matter halos compared, for instance, to galaxy-sized halos \citep{flores2007shape}. Analyses of simulation data \citep[e.g.,][]{cui2018three, Kuchner2020} indicate that cluster {halos} are well approximated by ellipsoids defined by three principal axes, with a tendency toward prolateness over oblateness. From the observational side, it has been demonstrated that galaxy clusters are not spherical objects based on evidence from the optical distribution of cluster galaxies \citep[e.g.,][]{carter1980morphology,binggeli1982shape,shin2018}, Sunyaev-Zel’dovich pressure maps \citep[e.g.,][]{sayers2011cluster}, strong and weak gravitational lensing \citep[e.g.,][]{oguri2009subaru,oguri2010direct,oguri2012combined}, and surface brightness maps in X-ray \citep[e.g.,][]{kawahara2010axis,lau2012constraining}. For a detailed review about the asphericity of galaxy clusters, see \cite{Limousin2013}. 

Several studies have shown that the elliptical shape of galaxy clusters is not fixed, but varies with redshift ($z$) and mass \citep[e.g.,][]{allgood2006shape,despali2014some,suto2016evolution,vega2017shape}, with more massive halos tending to be more elliptical than their lower-mass counterparts. Additionally, clusters connected to a larger number of filaments tend to be more elliptical, less dynamically relaxed, and accreting more mass \citep{gouin2021shape}. The latter, combined with the fact that filaments are the main channels through which galaxy clusters are fed, suggests that a useful approach to studying cluster accretion is to characterize their gravitational potential and examine how it correlates with the surrounding filaments. This can help to identify which regions of clusters are most affected by filamentary inflow and provide insights into their dynamical state.

One of the most direct ways to characterize the geometry and gravitational potential of galaxy clusters in observations, and consequently the mass accretion onto them, is through X-ray surface brightness maps. It is well established that galaxy clusters emit X-rays \citep{sarazin1986x}, with this radiation originating from the diffuse gas distributed within them, the so-called intra-cluster medium (ICM), which typically resides at temperatures of $10^7 - 10^8$ K. Unlike galaxies in clusters, the ICM is a collisional system that, as the gas falls into the cluster potential well, it converts its kinetic energy into thermal energy \citep{kravtsov2012formation} through the thermal bremsstrahlung process, the main cooling mechanism of the ICM \citep[e.g.,][]{sarazin1986x},  transforming the kinetic energy of the ions into X-ray radiation. 

One of the best environments to study the relationship between clusters and their surroundings is within superclusters. Superclusters can be considered as miniature Universes that contain all the large-scale structure components at the same time, representing the densest and most dynamically active environments where galaxies and their systems formed and evolve \citep{einasto2021}. Since superclusters are not virialized structures, matter within them continues to interact dynamically, flowing along gravitational potentials toward the most massive structures and giving rise to the formation of walls, filaments, groups, and clusters \citep{galarraga2020}. As was shown in \cite{tanaka2007huge}, the likelihood of finding filamentary structures is higher in superclusters. Following the supercluster classification by \cite{einasto2014sdss}, such structures can be identified regardless of whether the supercluster is of the filament-type or spider-type, with filaments appearing in linear configurations in the former and in more radial patterns in the latter. 

Given the above, the main goal of this work is to observationally determine, for the first time, whether there is a direct correlation between the projected orientation of optical filaments (as traced by galaxies) and the distribution of X-ray surface brightness in dozens of galaxy clusters that are part of two large nearby supercluster regions: the Shapley supercluster (SSC) and the Horologium-Reticulum supercluster (HRSC). We propose this measurement as a simple method to study the accretion process onto clusters through filaments.

Throughout this paper, $M_{\Delta}$ (where $\Delta \in \{200,500\}$)  refers
to the mass within $r_{\Delta}$, corresponding to the radius enclosing a
density $\Delta$ times the critical matter density of the Universe at
each redshift. Additionally, we assume cosmological parameters inferred by \cite{planck2020},  the most relevant of which are
the current expansion rate, $H_0 = 67.4$ km s$^{-1}$ Mpc$^{-1}$, and the
present-day matter density parameter, $\Omega_{m} = 0.315$.

 For simplicity of the analysis, even though the filamentary
   structure that we study is inherently highly inhomogeneous, we
   ignore the question of how to calculate an averaged cosmological
   expansion model \citep{Buch00scalav,Buch01scalav,
   rasanenFOCUS,KrasinskiHellaby,BCK11review,wiltshireFOCUS,
   GWdebunk15,Koksbang2024}
   and instead we adopt the usual assumption that filamentary structure
   dynamics are affected by the expansion, but the expansion itself is
   predetermined by the cosmological model parameters as
   a ``background'' unaffected by the filamentary structure.

The paper is organized as follows. In Sect.~\ref{s:sample}, we provide a brief overview of previous studies on the two superclusters considered in this work. In Sect.~\ref{s:data}, we describe the optical catalog used to identify the LSS in both superclusters, as well as the X-ray data employed to characterize the galaxy clusters within them. In Sect.~\ref{sec:method}, we present the methods used for LSS identification, the determination of cluster shapes and inclinations from X-ray data, the projected inclination of filaments, and the analysis of cluster–filament alignments. In Sect.~\ref{sec:results}, we present the main results. Finally, conclusions and a summary of this work are provided in Sect.~\ref{conclusions}.

\begin{figure*}
    \centering
    \includegraphics[width=.95\linewidth]{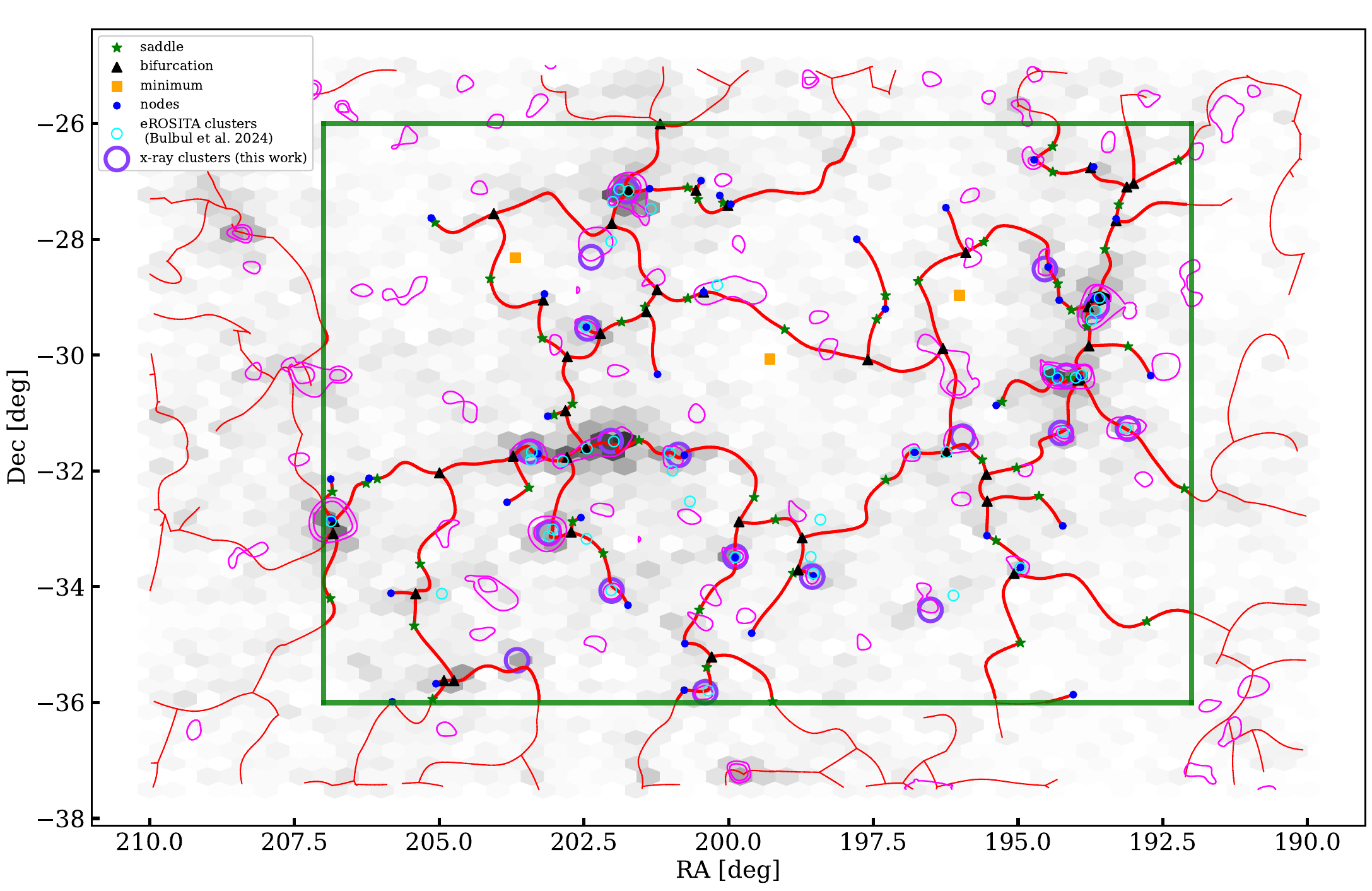}
    \caption{Large-scale structure identification using DisPerSE in two dimensions within the SSC. Greyscale hexagons show the density map of photometric members in the supercluster. Red lines indicate optically detected filaments, while different symbols correspond to critical points (see legend). Large-scale X-ray detections in the SSC area are in magenta contours. The cyan circles indicate clusters detected in the 0.2-2.3 keV as extended X-ray sources in the Shapley area and $z$ range considered for this work {\citep{Bulbul2024}}, while purple circles are the large-scale emission in the band 0.6-2.3 identified as galaxy clusters used in this work. Green rectangle indicates the CHANCES coverage in SSC and the area considered for the optical-X-ray comparative analysis shown in Sect.~\ref{sec:method}.}
    \label{fig:Shapley}
\end{figure*}

\begin{figure*}
    \centering
    \includegraphics[width=0.7\linewidth]{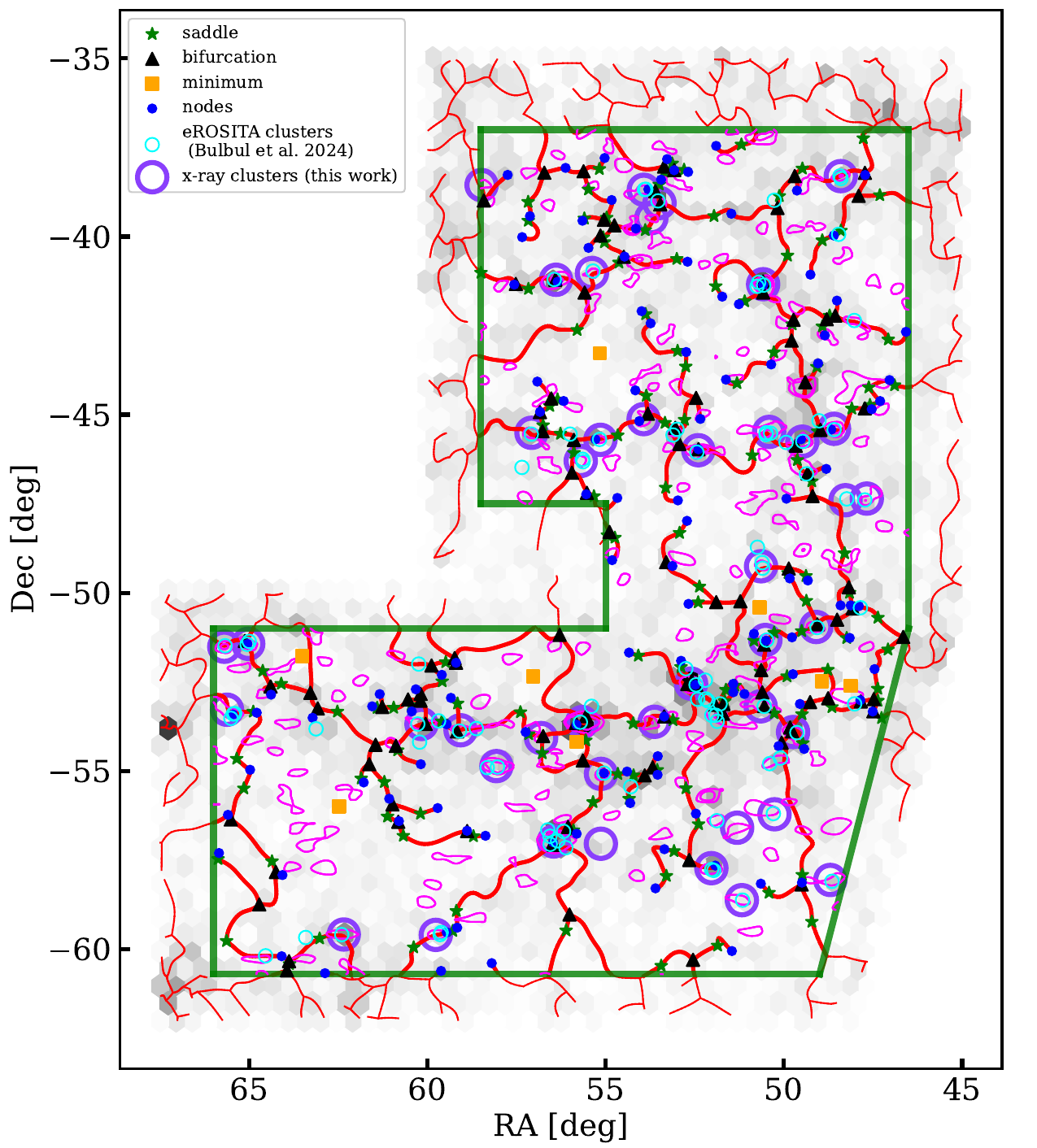}
    \caption{ Same as Fig.~\ref{fig:Shapley} but for HRSC. Green polygon indicates the area that CHANCES will observe.}
    \label{fig:Horologium}
\end{figure*}

\section{Overview of the target superclusters}\label{s:sample}

 The two large nearby superclusters studied in this paper were chosen because they contain dozens of clusters interconnected by filaments, and they have good optical and X-ray data available. In addition, they lie within the footprint of the extended ROentgen Survey with an Imaging Telescope Array (eROSITA) X-ray mission and the upcoming CHileAN Cluster galaxy Evolution Survey {\citep[CHANCES;][]{sifon2025}}, which will obtain over 500000 galaxy spectra in and around these two superclusters, as well as in 100 other clusters at $0 < z < 0.45$, using the 4 metre Multi-Object Spectrograph Telescope \citep[4MOST;][]{deJong2019Msngr}. The survey aims to enable a more detailed and deeper analysis of these structures. Our work serves as a precursor to CHANCES and contributes to validating its target selection strategy \citep{MendezHernandez2025}. In the following (Sect.~\ref{s:sample} and~\ref{s:data}), we describe the superclusters and the data available.

\subsection{The Shapley supercluster}
\label{subsec:SSC}
The SSC is one of the most massive and densest superclusters in the local Universe, encompassing 25 clusters in the redshift range $0.03 < z < 0.06$ over a $\sim 15 \times 10$\,deg$^2$ region \citep{Raychaudury89,Quintana1995,Ettori97,DeFilippis05,Proust06,Quintana2020}. Previous studies, {such as \cite{Quintana1995}, based on spectroscopy within and up to 7 deg around the center of the supercluster,} indicated that the SSC has a cigar-shape with the eastern side being the closest to us, showing a flattened geometry, which suggests that it is not spherical or virialized. The complex morphology of the SSC comprises a main body at
$cz\sim$15000\,km\,s$^{-1}$ together with walls and filaments of
galaxies connecting the three main systems of interacting clusters
(the A3558, A3528 and A3571 complexes) as well as a foreground
structure connecting the SSC to the Hydra-Centaurus supercluster
($cz\sim$4000\,km\,s$^{-1}$). 

At the heart of the SSC is the high-density SSC core (Dec $\sim -31$\,deg) at $z = 0.048$, comprising three Abell clusters and two poor clusters, forming an elongated structure 2 degrees ($\sim 7$\,Mpc) in extent filled with hot gas \citep{Planck13}. Past dynamical \citep[e.g.,][]{Bardelli94,Ragone06} and X-ray \citep[e.g.,][]{KB99,Ettori00,Finoguenov04} studies revealed cluster-cluster interactions in the core region that is characterized by a complex dynamical state with several subcondensations and a diffuse filamentary X-ray emission across the whole region. More recently, \citet{Venturi2022} detected a {low-brightness} intercluster diffuse emission at 1.28\,GHz in the SSC core and interpreted it as a radio signature of a minor merger. Using the Shapley Supercluster Survey \citep[ShaSS,][]{Merluzzi2015}{,} \citet{Haines2018a} obtained a detailed map of the region around the supercluster core, which includes 11 SSC clusters. They showed that the 11 systems are all {interconnected} and lie within a coherent sheet of galaxies that fills the entire survey region without gaps \citep[see also][]{Reisenegger2000}. This study also revealed a stream of galaxies from the northern cluster A3559 to the core. 

We notice that the dynamical analysis of \citet{Haines2018a} is the only study based on a spectroscopic catalog which is uniformly complete\footnote{The ShaSS spectroscopic catalog is 95 percent complete at $i$=18 \citep[see][for details]{Haines2018a}.} across the considered region of the SSC ($\sim 21$ deg$^2$ around the SSC core). However, this work is limited to the central region of a very complex structure. Other studies \citep[e.g.,][]{Proust06,Quintana2020,Aghanim2024} traced the whole supercluster structure covering up to 300\,deg$^2$, agreeing on the map of the filaments as well as with the findings of \citet{Haines2018a} for the SSC core. However, these works rely on the photometry from the SuperCOSMOS scans of photographic Schmidt plates\footnote{\url {http://www-wfau.roe.ac.uk/sss/index.html}}. They are based on collections of spectroscopic redshifts that, as expected, are highly complete ($>$80 percent) at the bright end of the galaxy luminosity function ($r\leq 15$). Still, their completeness drops to less than 20 {percent} at $r=17.5$ \citep[see Fig. 3 of][]{Quintana2020} implying that the filaments turn out to be traced mostly by a few bright supercluster galaxies. CHANCES will go a step further, not only providing an accurate map of the filaments and infalling groups, but in particular,  mapping the evolution of the supercluster members well outside the cluster cores and down to $r=20.4$ - about M$^\star+5$ at the SSC redshift.

\subsection{The Horologium-Reticulum supercluster}

The HRSC is the {second-largest} mass concentration in the local Universe \citep{zucca1993all,johnston2008radio}, {surpassed only} by the SSC. The HRSC contains more than $20$ optically identified galaxy clusters \citep{einasto2002optical} in the redshift range $0.04<z<0.08$ within a region of $\sim$ 150 deg$^{2}$ on the sky \citep{fleenor2006redshifts}, and is composed of at least two major filaments \citep{einasto2003campanas}. HRSC can be separated into two main groups of galaxy clusters \citep{einasto2003campanas}, the northern clusters ($-48 \text{ deg} <$ Dec $<-43\text{ deg}$) at $z \sim 0.07$ and the southern clusters ($-57\text{ deg}<$ Dec $<-51\text{ deg}$) at $z\sim 0.06$ \citep{fleenor2006redshifts}. The HRSC center is defined by the double cluster complex A3125/ A3128 (RA $\sim 52\text{ deg}$, Dec $\sim -53\text{ deg}$) \citep{rose2002multiple}. At $\sim$ 2.5$\text{ deg}$ ($\sim$ 10 $h^{-1}$ Mpc) east of the A3125/A3128 complex is located A3158, a compact cluster of richness class two \citep{quintana1979detailed}. These three clusters have long been known to be linked by a ``bridge'' of galaxies \citep{lucey1983horologium} within the larger HRSC supercluster structure. The multiwavelength analysis performed by \cite{rose2002multiple} in A3125/A3128 uncovered rapidly infalling groups and filaments accelerated by the HR potential, suggesting evolving substructures across various mass scales.

Compared to the SSC, the dynamical state of the HRSC environment outside rich systems has not been extensively studied. \cite{fleenor2006redshifts} conducted one of the first studies focusing on the intercluster regions in HRSC by performing wide-field spectroscopy with the Dual Beam Spectrograph. From this study has been revealed a main concentration of intercluster galaxies covering $cz \sim$ 17000 - 22500 km s$^{-1}$ that consists of two major components in redshift space, separated by
2500 km s$^{-1}$ (35 Mpc), each with a similar inclination in the position-redshift space at the same position angle. 
 
In summary, their high density and gravitational influence make SSC and HRSC significant sites for studying large-scale structure formation and the impact of the environment on galaxy clusters{,} as well as galaxy evolution.

\section{Data}\label{s:data}

To characterize LSS in supercluster regions and identify the filaments connecting clusters within them, we use optical galaxy data from superclusters selected in the context of target selection for the CHANCES 4MOST survey (Sect.~\ref{subsec:chancesTS}). These structures are then compared with the X-ray emission observed by eROSITA, as described in Sect.~\ref{subsec:xrays}.

\subsection{Optical data from the CHANCES target catalog}
\label{subsec:chancesTS}

While SSC and HRSC have been studied previously, no deep and wide homogeneous spectroscopic datasets are yet available to examine them in detail. This has motivated CHANCES to include these two superclusters in their survey. CHANCES will target galaxies in these regions down to $r=20.4$, along with galaxies in other 100 clusters at $z<0.45$ to be observed with the 4MOST spectrograph. 
CHANCES is split in three sub-surveys:
CHANCES Low-z ($z<0.07$), CHANCES Evolution ($0.07<z<0.45$) and CHANCES Circumstellar Galactic Medium at $z>0.35$ \citep[see][for details]{Haines2023}.  The Low-z CHANCES sub-survey targets galaxies brighter than $r=20.4$ in 50 $z<0.07$ clusters out to $5\times r_{200}$ and the two large supercluster regions (SSC, HRSC){,} which are the focus of this paper. Cluster selection and properties as well as  the definition of the superclusters regions are detailed in {\cite{sifon2025}}.

Galaxy target selection within the superclusters (and CHANCES clusters more broadly) was guided by a combination of photometric redshifts to efficiently identify likely cluster members and exclude outliers. As explained in detail in \cite{MendezHernandez2025}, to select targets for CHANCES Low-z sub-survey\footnote{The CHANCES-low-z sub-surveys were compiled using the Efficiently Extracting Cluster candidate members
with Homogeneity using Optical colour and photometric-z \citep[\texttt{EECHOz};][]{EECHOz} code, available at  \url{https://github.com/4MOST-CHANCES/CHANCES-EECHOz-Low-redshift-TargetSelection}.}, we used both publicly available photometric redshifts ($z_{\text{phot}}$) from Legacy Survey DR10 \citep{zhou21}, as well as proprietary $z_{\text{phot}}$ from S-PLUS. For fainter galaxies ($r > 18.5$), we use a combination of S-PLUS custom-derived photometric redshifts based on the Legacy Survey to minimize bias against faint red galaxies ({Teixeira} et al., in prep). In this paper, since the aim is to characterize the LSS minimizing any possible biases, we focus only on bright targets ($r < 18.5$) where photometric redshifts are most reliable. We also focus exclusively on Legacy Survey DR10 photometric redshifts (and exclude S-PLUS) to ensure homogeneity, as all the area studied is covered homogeneously by Legacy Survey.

Additionally, as in \cite{MendezHernandez2025}, available spectroscopic redshifts from the literature were used to confirm supercluster members and remove outliers, {and to compute the completeness and purity of the target catalog. Catalog completeness was defined as the fraction of previously known spectroscopic members of the superclusters that are successfully retrieved through the photometric target selection, while purity was defined as the fraction of the selected photometric supercluster candidates that have been spectroscopically confirmed as members. For SSC, completeness and purity are 0.93 and 0.70, respectively, and for HRSC they are 0.91 and 0.94.} 

The $z_{\text{phot}}$ ranges of bright CHANCES low-z sub-survey targets for SSC and HRSC — $0.027<z_{\text{phot}}<0.073$ and $0.037<z_{\text{phot}}<0.083$, respectively — were defined to include both photometric member galaxies and the clusters associated with each supercluster. Note that these ranges are wider than those assumed in previous works (for example, ${0.035<z<0.058}$ {in} \cite{Haines2018a} and ${0.03<z<0.06}$ {in} \cite{Quintana2020} for SSC, and ${0.04<z<0.07}$ {in} \cite{rose2002multiple} for {HRSC}). This discrepancy arises from our use of $z_{\text{phot}}$ for membership determination, which introduces larger uncertainties compared to spectroscopic membership.

The CHANCES coverage of the SSC consists of 128 deg$^2$ plus the remaining area within $5r_{200}$ of A3571 {\citep[see][]{sifon2025}}. 
The SSC region includes $\sim 25$ galaxy clusters at $0.037<z<0.056$ \citep[e.g.,][]{Haines2018a}, with A3558, A3562, and A3528 as prominent components. Other low-$z$ clusters overlap the superclusters in the sky (namely, A3565 and A3574 with Shapley), but they are at noticeably different redshifts, and target overlap is minimal. For the case of HRSC, CHANCES covers an area of $\sim 225$ deg$^{2}$, containing the HRSC center (RA $\sim 52$ deg, Dec $\sim -53$ deg) plus the marginally overlapping $5r_{200}$ area around A3266, at a mean $z=0.06$. Similar to SSC, just at the northern border of the HRSC there is a partial area overlap with a lower redshift Fornax cluster, whose contamination is removed using redshift information.

In this work, we considered areas $\sim 1-2$ degrees wider than the CHANCES coverage of the superclusters to characterize the LSS without introducing edge effects within the CHANCES fields. The studied areas can be appreciated in Fig.~\ref{fig:Shapley} and~\ref{fig:Horologium} along with the CHANCES fields (green).

\subsection{X-ray data from eROSITA}
\label{subsec:xrays}

The CHANCES survey has a great synergy with eROSITA, complementing the optical information with X-rays to better characterize large-scale structure. The two superclusters studied in this paper are fully covered by eROSITA. 

 eROSITA is a German telescope on board the Spectrum-Roentgen-Gamma satellite, which has performed an all-sky X-ray survey, and has deeper sensitivity compared to the previous ROSAT All-Sky Survey, with a spatial resolution of $30^{\prime\prime}$ \citep{Predehl21}.  
The publicly released year 1 eROSITA-DE data (hereinafter eRASS1) cover half of the sky. We used both the 0.6-2.3 keV and 0.2-2.3 keV band images of eROSITA, together with the officially released exposure and background files\footnote{\url {https://erosita.mpe.mpg.de/dr1}}. We perform a wavelet image decomposition \cite[][\footnote{\url {https://github.com/avikhlinin/wvdecomp}}]{vikhlinin98} to separate the unresolved emission from the cluster emission. Further details on the wavelet decomposition method can be found in \cite{vikhlinin98}. The sources detected on spatial scales up to 32 arcseconds are excised from the flux extraction, a procedure similar to other published eROSITA analyses \citep{reiprich25}. For the redshifts of this study, we benefit from the image reconstructions on scales up to 16 arcminutes. Larger scales require reconsidering the background estimates obtained using in-field estimates after source removal. For the science of this study, these spatial scales are not required.

In our work, we used the 0.6 - 2.3 keV band for the analysis of large-scale emission. This is because the 0.2--2.3 keV band is confusion-limited on scales of 16 arcminutes and has a higher background, due to Galactic foreground emission, and in addition is subject to solar leaks, which lead to a temporal increase in the level of the background below 0.6 keV. In estimating the shape of the emission{,} we used the surface brightness in units of counts/s/pixel, in which we used the binned pixels of 32 arcseconds on a side.  

We have run Source Extractor \citep{sextractor1996} to characterize the sources detected on scales of 2-4 and 8-16 arcminutes, storing the positional angle together with values of major and minor axes. To select the sources associated with the superclusters and to remove clusters at other redshifts, we run the identification of sources using version 8 (Python) of the red-sequence Matched-filter Probabilistic Percolation cluster-finding algorithm code \citep[redMaPPer,][]{Rykoff2014} in scan mode \citep[e.g. as in][]{IderChitham2020,kluge2024srg}, which for the location of our sources utilized the 10th Data Release of the DESI Legacy Imaging Surveys \citep[LS-DR10;][]{dey2019}. To complete the identification of sources, we added the eROSITA sources identified with the 2MRS group catalog \citep{Tempel18}, which has better sensitivity towards low-mass systems at $z<0.04$, important for the SSC.
Tables~\ref{tab:x_ray_table} and~\ref{tab:x_ray_table_HRSC} show the main properties of the detected X-ray clusters in SSC and HRSC, respectively.  We present the X-ray luminosity in the rest-frame 0.1-2.4 keV, applying band and K-corrections computed using the eRASS1 release of eROSITA responses. We refer to \citet{Isabel25} for the weak lensing calibration of eRASS1 luminosities in the 0.1-2.4 keV band and a comparison to previous scaling relations. We report masses and virial radii computed using those scaling relations and X-ray luminosities.

\section{Method}\label{sec:method}

Our goal is to compare the shape and alignments of individual clusters to the filaments that connect them in supercluster regions. To do that, we first identify the filament network (Sect.~\ref{subsec:optical_identification}), before determining the inclination angles of filaments to the galaxy clusters (Sect.~\ref{subsec:filaments_PA}). Finally, we fold in the information of shapes of clusters inferred from X-ray emission (Sect.~\ref{subsec:x-ray_shape})

\subsection{Identification of cosmic filaments}
\label{subsec:optical_identification}

Filament detection in both superclusters was performed using the Discrete Persistence Structures Extractor \citep[DisPerSE\footnote{\url{https://www2.iap.fr/users/sousbie/disperse.html}};][]{Sousbie2011}, a topological tool based on discrete Morse theory and persistence theory. DisPerSE has been widely applied to identify filaments, nodes, walls, and voids from discrete source distributions, both in simulations \citep[e.g.,][]{galarraga2020, Kuchner2020} and in observational data \citep[e.g.,][]{Bonjean2020, smith2023bcg}. 

In short, DisPerSE uses the point distribution, in our case the galaxy coordinates, to reconstruct the area of the sky that we are studying as cells, edges, and vertices. To estimate the density field of this distribution, it employs the Delaunay tessellation, where the density around each vertex of the Delaunay complex is calculated using the Delaunay Tessellation Field Estimator \citep[][]{schaap2000continuous,cautun2011dtfe}. 
To extract filaments and nodes from this density field, DisPerSE identifies the critical points where the gradient vanishes --such as maxima, minima, and saddle points – and connects them along ridges. The filamentary structure is traced by field lines tangent to the gradient at each point. These filaments are identified by computing small segments that connect topological saddle points to maxima (nodes), forming a continuous skeleton of the cosmic web. In this network, filaments are represented as arcs connecting critical points: maxima, which are of third order in three dimensions and second order in two dimensions, and saddle points, which are of second or first order in three dimensions and of first order in two dimensions.

The significance of each connection is quantified by the persistence, a measure of the significance of topological connections between critical points, and is defined as the density contrast between a pair of critical points. The persistence is typically expressed in units of its standard deviation $\sigma$. A higher value of $\sigma$ allows us to filter and remove noisy structures.

In our work, the filaments were detected in two dimensions, using as input the positions of the photometric members on the sky, considering for each supercluster just one redshift slice that encompasses the total redshift range mentioned in Sect.~\ref{subsec:chancesTS}. The Delaunay tessellation was done assuming smooth boundary conditions. To extract filaments and critical points, we assumed a persistence threshold of $3\sigma$, and a smoothness level of the skeleton of 20. Fig.~\ref{fig:Shapley} and~\ref{fig:Horologium} show the LSS identification for SSC and HRSC, respectively. Different symbols represent different critical points given by DisPerSE. At the same time, we have included in the two figures the large-scale X-ray detections (magenta contours) from our own data reduction (see Sect.~\ref{subsec:xrays}) in the SSC and HRSC area, as well as the clusters from the {eRASS1 Galaxy groups and clusters primary catalog {\citep[cyan circles;][]{Bulbul2024}} that lie within the redshift range assumed for photometric members. Note that the resolution on the density map of photometric members (grayscale density map) used in the figures is too low to see all the clusters from the public catalog. Additionally, in Fig.~\ref{fig:Shapley} and~\ref{fig:Horologium}, we have included the X-ray catalog of large-scale sources used in this work as purple circles. Note that, compared to the public catalog, the sources in our catalog are extended, and in some cases, multiple sources listed separately in the public catalog correspond to a single source in our case. 

To validate our method for selecting supercluster members and identifying filaments, we compared the filamentary structures detected using photometric data with those obtained in a smaller region uniformly covered by spectroscopic observations from the ShaSS survey (see Sect.~\ref{subsec:SSC}), complemented by a compilation of additional spectroscopic data in the same ShaSS area from \cite{Quintana1995}, \cite{Quintana1997}, \cite{Drinkwater2004}, \cite{Proust06}, \cite{jones2009}, and \cite{Quintana2020}, to include the majority of spectroscopically confirmed sources in the region. A match of the sky positions was performed in order to remove sources with duplicates across the different spectroscopic catalogs. Filaments from the spectroscopic catalog were detected in two dimensions, like those from the photometric catalog, using a galaxy sample within the redshift range adopted by \cite{Quintana2020} ($0.03 < z < 0.06$), providing a more accurate membership of galaxies in the SSC core, while for the filaments extracted from the photometric sample we adopted the redshift range for the SSC described in Sect.~\ref{subsec:chancesTS}. The left panel of Fig.~\ref{fig:filaments_comparison_PDF_CDF} shows in orange the filaments detected from the photometric catalog in that area and in green the filaments obtained using the spectroscopic data. Note that the orange filaments in this Fig. are not the same as the red filaments in Fig.~\ref{fig:Shapley} due to the different area in which we ran DisPerSE. However, the main structures in the SSC core prevail. When comparing the spectroscopically detected filaments with the photometric ones in the same area, we find that qualitatively, the two networks agree well, except for a photometric filament detected at RA $\sim$ 201$\ \text{deg}$ that is not detected on the spectral sample. We notice that \cite{Higuchi2020}, using ShaSS data, measured a mass peak at RA=202.1 deg, Dec=-32.7 deg, identified as a background cluster at $z=0.17$. The cluster position matches that of the pronounced curve of the photometric redshift filament. This background cluster, together with other structures behind the supercluster \citep[see Fig. 7 of][]{Haines2018a}, may contribute to tracing the filaments, considering the wider redshift range adopted for the photo-z and their larger errors with respect to the spectroscopic redshifts. 

To quantify the similarities (or discrepancies) between the LSS skeletons obtained from the photometric and spectroscopic samples, we followed the method introduced in \cite{Sousbie2011} and used in \cite{sarron2019pre}, \cite{laigle2018cosmos2015}{,} and \cite{Kuchner2020}, which provides an indicator of the reliability of filament extraction using two-dimensional photometric and spectroscopic data. For this purpose, we measured the projected distances between the photometric and spectroscopic skeletons to obtain the differential distribution (PDF). The projected distances were computed between each segment point on the photometric network and the nearest segment point on the spectroscopic network. The right panel of Fig.~\ref{fig:filaments_comparison_PDF_CDF} shows the resulting PDF obtained by comparing photometric to spectroscopic (orange curve) and spectroscopic to photometric (green curve) filaments, from which we obtained median values of $0.36$ Mpc and $0.29$ Mpc, respectively. Given the common assumption that filament thickness is of the order $\sim 0.7 \ -\  1$ Mpc \citep[e.g.,][]{colberg2005intercluster,tempel2014detecting,martinez2016galaxies,kooistra2019detecting,sarron2019pre,Kuchner2020}, our median values suggest a typical offset smaller than the characteristic filament thickness.

\begin{figure*}[htbp]
    \centering
    \begin{minipage}{0.48\linewidth}
        \centering
        \includegraphics[width=\linewidth]{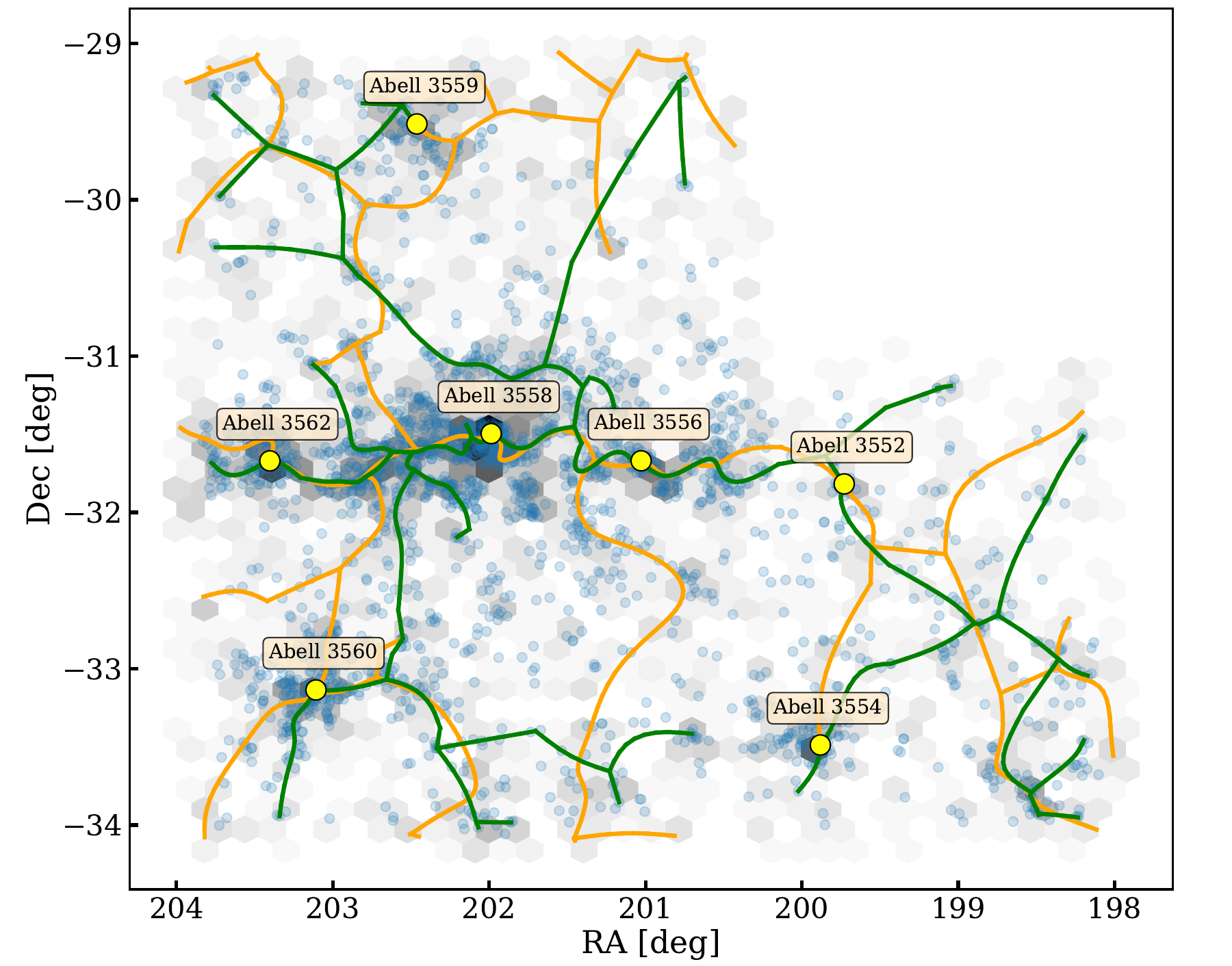}
    \end{minipage}
    \hfill
    \begin{minipage}{0.48\linewidth}
        \centering
        \includegraphics[width=\linewidth]{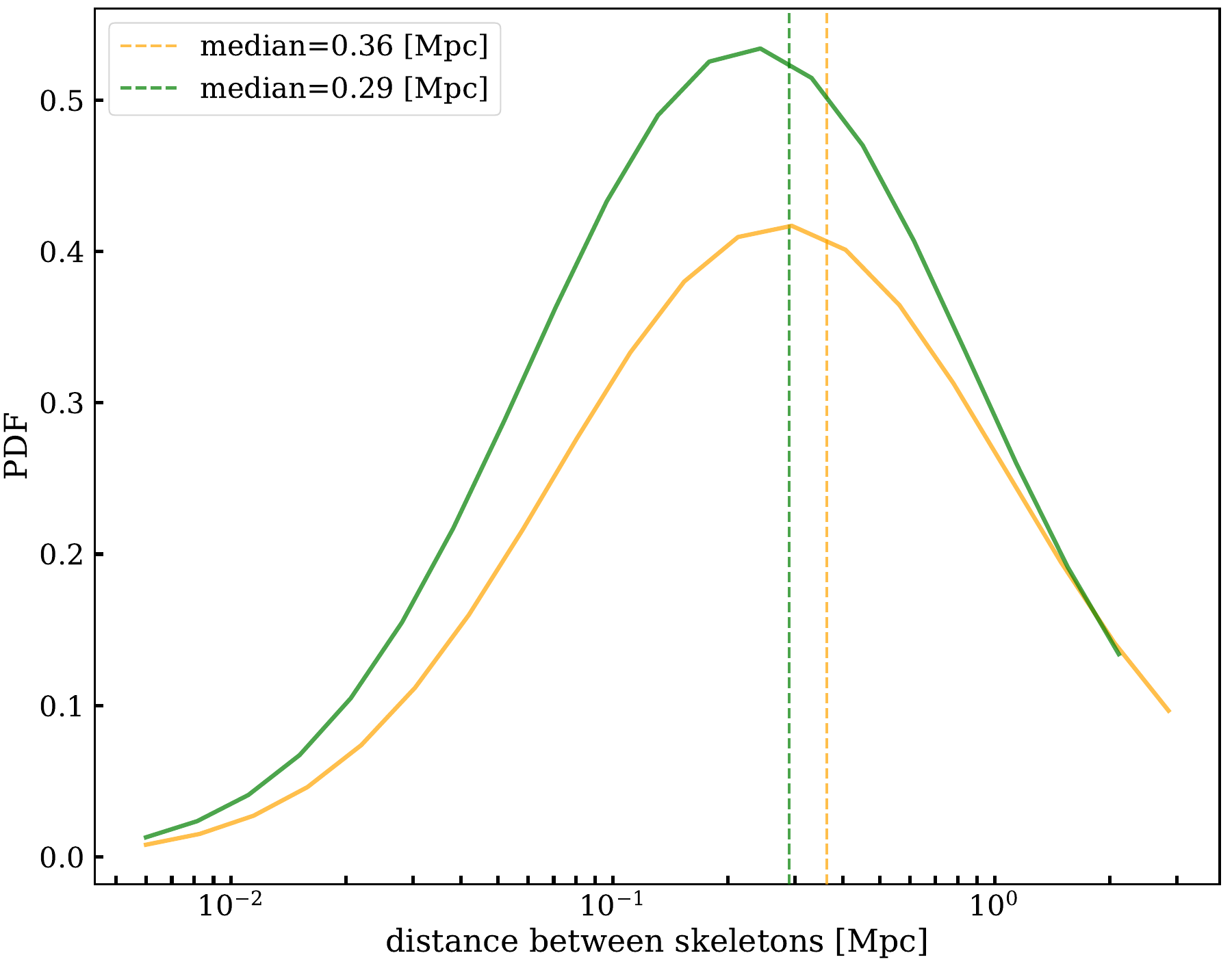}
    \end{minipage}
    \caption{\textit{Left:} Two-dimensional reconstruction of the filament network using the spectroscopic sample (green segments) and photometric members (red segments) in the SSC core. The gray map indicates the density of photometric members in the area, while blue circles are the spectral members. Orange circles indicate known X-ray clusters in the SSC \citep{Haines2018a}. \textit{Right:} Probability distribution of the distances between skeletons of filament networks from photometric to spectroscopic (orange curve) and spectroscopic to photometric (green curve) members within the ShaSS area. Vertical dashed orange and green lines show the median distance between skeletons obtained from the two comparisons.}
    
    \label{fig:filaments_comparison_PDF_CDF}
\end{figure*}

\subsection{Characterization of cosmic filaments}
\label{subsec:filaments_PA}

We ultimately want to compare the location and projected orientation of filaments with the shape of the clusters. To do this, we first create catalogs of optical filaments connected to X-ray clusters, adopting the definition used in \cite{Kuchner2020}. In that work, a filament is defined as a segment that is connected to a cluster and extends beyond a sphere of  1 $r_{200}$ (in three-dimensional space) around the center of the main dark matter halo. Since our work was carried out using two-dimensional data (positions projected in the plane of the sky), the corresponding set of filaments connected to a cluster is those exiting a circumference projected on the sky of $1 r_{200}$ centered on the position of the X-ray peak. To compute the projected inclination, we use the filament segment located between 0.8 and 1.2 times the $r_{200}$ of the corresponding X-ray cluster. This range is chosen to approximate a straight section of the filament, ensuring a more reliable inclination estimate. In addition, we also compute projected filament inclinations at 1.4–1.8 and 3.0–3.4 times the $r_{200}$ to investigate how the alignment between filaments and clusters varies with distance from the cluster center.

Then, to determine the average projected inclinations of filaments connected to galaxy clusters, we used the Probabilistic model-based Hough Transform technique \citep[PHT,][]{tino2011searching}, a Bayesian method to determine alignments in data points based on the original idea of the Hough Transform \citep{hough1962method}. If $N$ is the total number of segment points of a filament, a Gaussian noise model with covariance $\Sigma_{\alpha,r}$\footnote{For this work, we assumed $    \Sigma_{\alpha,r}= \begin{bmatrix}
\sigma_{\alpha}^{2} & 0 \\
0 & \sigma_r^{2} \end{bmatrix}$, with $\sigma_{r}=0.35\text{ deg}$ and  $\sigma_{\alpha}=20.5\text{ deg}$} is assumed for each segment point. 
The model assumed in this work is given by

\begin{equation}
\begin{aligned}
    p({x}|\alpha,r)=\frac{1}{2\pi|\Sigma_{\alpha,r}|}\exp{\left[-\dfrac{1}{2}({x}^T - (r\cdot \cos\alpha,r\cdot\sin\alpha))\right]} \\ \Sigma_{\alpha,r}^{-1} ({x} - (r\cdot \cos\alpha,r\cdot\sin\alpha)^T) .
\end{aligned}
\label{eq:noise_model}
\end{equation}

From this, PHT determines the probability density for each point ${x}$ of belonging to a straight-line model defined by the parameter pair $(\alpha, r)$ — the inclination angle and the perpendicular distance from the origin to the line in polar coordinates, respectively. In our case, the model corresponds to a straight line passing through the origin, and includes a neighborhood region around the origin. In our case, the origin corresponds to each point along the filament segment over which we iterate during the computation. This origin changes from point to point. However, the distance to neighboring points — computed in two dimensions on the sky as an angular separation — remains fixed. In this work, we adopt a constant neighbor distance of 1.0 deg, in order to include the largest possible number of points in the calculation of the projected inclination of the filament segment, even near the segment ends where fewer neighboring points are available. Note that, given the observational noise $\Sigma_{\alpha,r}$, points closer to the origin matter less than those farther away, so it is important to adopt a neighbor distance that accounts for the overall segment inclinations. After that, we compute the marginal posterior over the angle parameter by integrating out $r$, conditioned on the observation $\mathbf{x}$. 

Finally, the Hough accumulator is obtained by summing all the marginalized probabilities to obtain the set of inclination angles $\mathcal{A}=\{\alpha_{1}^{max}, \alpha_{2}^{max}, ..., \alpha_{N}^{max} \} \in [0 \text{ deg} ,180\text{ deg}]$ of the line to which the greatest number of points belongs, i.e., the inclination angles where there is the greatest number of aligned points in all the $N$ segment points. Further details on the PHT method can be found in \cite{tino2011searching}. 

The overall projected inclination angle, $\phi_{fil}$, of a filament connected to a cluster is computed considering the vector average, which is determined from the set of complex components $X=\{ \cos{(2\alpha_{i}^{max})} \ | \ i = 1,2,..., N \}$ and $Y=\{ \sin{(2\alpha_{i}^{max})} \ | \ i = 1,2,..., N \}$. Thus, $\phi_{fil}$ comes from the following equation:

\begin{equation}
    \phi_{fil}=\frac{\arctan(\overline{Y}, \overline{X})}{2} \mbox{,}
    \label{eq:arctan2}
\end{equation}

 \noindent where $\overline{X}$ and $\overline{Y}$ are the mean values of $X$ and $Y$, respectively.

 Fig.~\ref{fig:PHT_cartoon} illustrates how filaments connected to X-ray clusters are defined, as well as how their projected inclinations are computed using the PHT technique. Note that the spacing between the segment points is not uniform, so in some cases, defining segments connected to X-ray clusters based on a projected distance criterion between 0.8 and 1.2 $r_{200}$ resulted in only one point per segment, making it impossible to determine a projected filament inclination. In those cases, the nearest adjacent point was also included in addition to the one within 0.8 to 1.2 $r_{200}$, in order to be able to define the projected inclination of the segment. 

\begin{figure}
    \centering
    \includegraphics[width=0.9\linewidth]{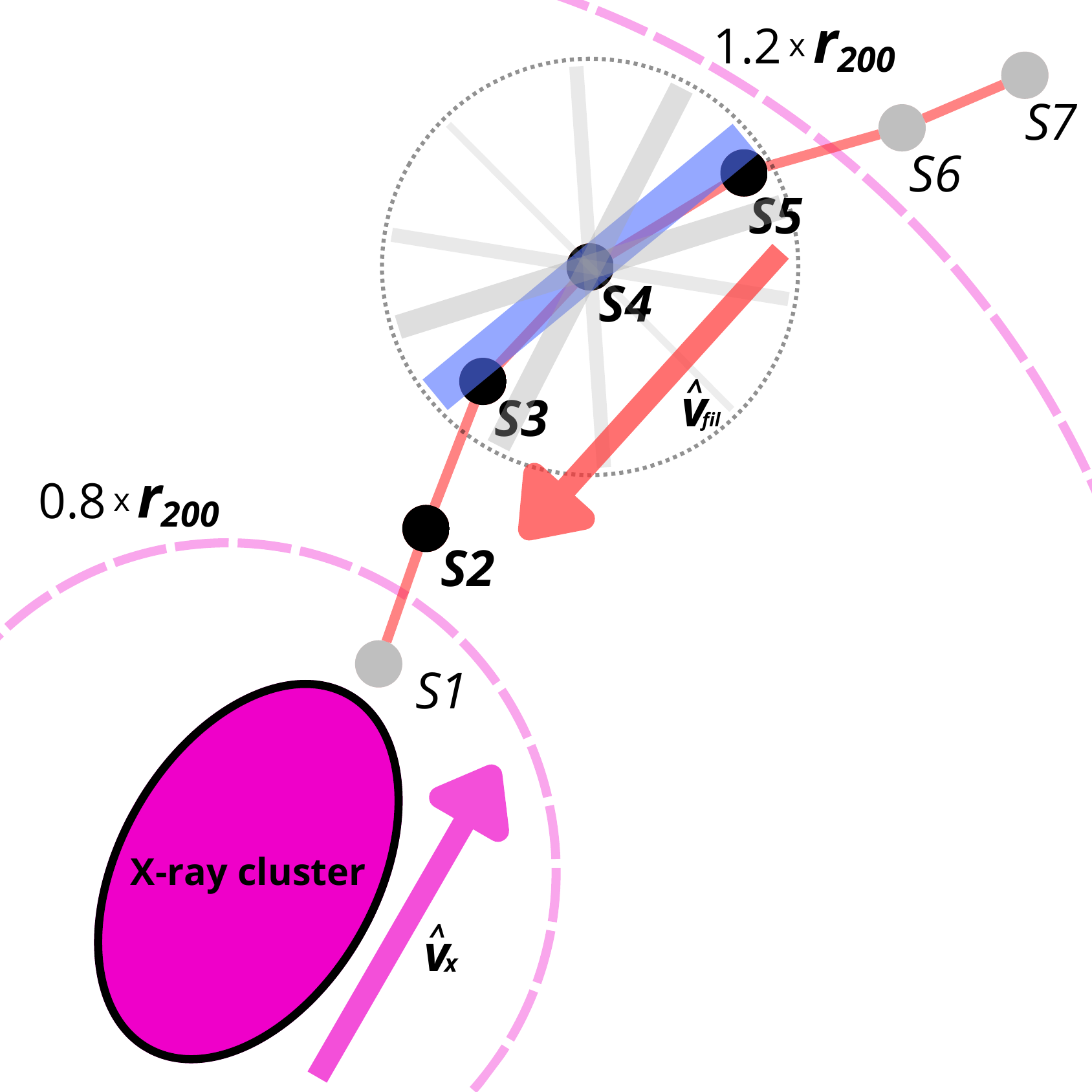}
     \caption{Schematic representation of the definition of filament segments connected to the X-ray cluster and the computation of their projected inclinations using the PHT technique. Red circles indicate the segment points $[S1,S2,…,S7]$ that are part of the detected filament network. Dashed magenta lines represent 0.8 and 1.2 times the $r_{200}$ around the X-ray peak, respectively. Dark red segment points $[S2,S3,S4,S5]$, located between the dashed curves, are used to compute the mean projected inclination of the filament segment connected to the galaxy cluster (magenta ellipse). The PHT operator is centered at the position of $S4$, with each line representing a possible local alignment angle of the segment points, based on the positions of the points within the dashed circumference. Line thickness corresponds to the probability of alignment at that specific angle, with the thickest blue line denoting the most probable inclination angle at $S4$. The same procedure is applied at $S2$, $S3$ and $S5$ to ultimately compute the mean projected alignment angle using Eq.~\ref{eq:arctan2}, represented as the red vector $\hat{v}_{fill}$. The inclination angle of the X-ray cluster is represented as the magenta vector $\hat{v}_{x}.$}
    \label{fig:PHT_cartoon}
\end{figure}

\subsection{Shapes of clusters inferred from X-ray emission}
\label{subsec:x-ray_shape}

Cluster shapes were characterized with the shape of the X-ray emission, which, for simplicity, was assumed to be ellipsoidal.  
Major and minor axes and the positional angle estimated by Source Extractor are intended to describe the detected object as an elliptical shape. $a$ and $b$ are the lengths of the semimajor and semiminor axes, respectively. More precisely, they represent the maximum and minimum spatial dispersion of the object profile along any direction. $\phi_x$ is the position angle of the $a$ axis relative to the first image axis, which in eRASS1 mosaics is RA with negative delta. It is counted positive in the direction of the second axis (which is Dec). By definition, $\phi_x$ is the position angle for which the major axis is maximized 
\footnote{\url{https://sextractor.readthedocs.io/en/latest/Position.html}}. Only wavelet images with unresolved spatial scales removed were used in our analysis. We provide the X-ray cluster characterization on both 2--4 and 8--16 arcminute scales and use the barycenter of X-ray emission defined by the corresponding scales.

\section{Results}\label{sec:results}

We have identified the LSS in both superclusters that are part of CHANCES, considering as tracers of the cosmic filaments the two-dimensional distribution of photometric members based on the target selection strategy described in Sect.~\ref{subsec:chancesTS}. Additionally, we have constructed a catalog of clusters detected on X-ray by eROSITA, comprising a total of 58 X-ray clusters, in a $z$ range of $0.027\leq z\leq0.073$ for the SSC, and $0.037 \leq z \leq 0.083$ for the HRSC, with estimated optical richnesses $\lambda$\footnote{The richness $\lambda$ is defined as the membership probabilities multiplied by the scaling factor $S$ \citep[see][for details]{kluge2024srg}.} ranging between $ 5<\lambda<90$. Of the total number of X-ray clusters, 19 are part of SSC, while the remaining 43 belong to {HRSC}. We have also created a catalog of filaments connected to these clusters. We determined and compared the inclinations of the clusters based on X-ray emission and their connected main filaments. In this section, we report the main results obtained from these analyses.

\subsection{X-ray clusters in the identified LSS}
\label{subsec:matches}

Since galaxy clusters are interconnected by cosmic filaments, it is natural to expect X-ray emission from clusters associated with nearby filaments. To compare the projected distribution of our detected filaments and X-ray clusters, as well as quantify their similarities (or discrepancies), we performed a cross-match between the X-ray clusters and the detected filaments, assuming the $r_{200}$ of each X-ray cluster as the threshold distance (projected on the sky) for the match. From this exercise, we found that 15 ($80 \%$) X-ray sources in the SSC are connected to at least one filament, while 48 ($ 78\%$) satisfy this condition in the HRSC. 

Based on the idea that clusters are located at filament intersections, we have considered both maxima and bifurcation points from DisPerSE as nodes. Under this assumption, we find that 63\% of the clusters detected in X-rays within the SSC are also detected in the optical (as node or bifurcation points detected by DisPerSE), while for HRSC this percentage raises to $67\%$. Of course, these fractions depend on the richness threshold adopted for X-ray clusters, the tolerance of positions in the sky, as well as the persistence level assumed for the structures detected on the projected distribution of photometric members.  

Taking the above into account, if both superclusters are considered, we find that $82\%$ of the X-ray clusters are connected to filaments. Additionally, $66 \%$ of the clusters were successfully detected based on the distribution of photometric members in the optical, a percentage similar to that obtained when considering only X-ray clusters connected to filaments.

\subsection{Typical cluster spacing in superclusters and field}
\label{subsec:min_separation}

In addition to comparing the filaments and nodes from the optical with the clusters in X-rays, we computed the typical projected separation between closest clusters in SSC and HRSC, in order to determine whether there is a significant difference compared to the average minimum separation of isolated clusters. As this value depends on the richness $\lambda$ threshold of the considered clusters, for this exercise, we just considered X-ray clusters with $\lambda>10$.

We obtained that the median minimum separation among X-ray clusters in the SSC is $\sim 4$ Mpc, while in the HRSC this separation is $\sim 4.6$ Mpc. We note that these values correspond to lower limits, as we are considering only the projected positions of clusters on the sky, rather than the three-dimensional distribution of clusters within the superclusters. For the same reason, the smaller value for the SSC may be due to the inclination of the line-of-sight axis with respect to us. As a control sample, we used the eRASS1 catalog of optical identifications of eRASS1 sources that are galaxy clusters from \cite{kluge2024srg} to estimate the average cluster separation. We considered the clusters in the redshift range $0.08<z<0.15$, similar as the redshift range considered for the SSC and HRSC, with $\lambda>10$. We used the clusters located in the deepest part of eRASS1 area 45 deg $<$RA$<$ 90 deg, -60 deg $<$Dec$<-25\text{ deg}$ to compute the distances to the nearest clusters in the full catalog with a redshift difference below 0.02. The median projected separation of the clusters was 8.3 Mpc. The value obtained for isolated clusters is $\sim 1.8$ times higher than the mean minimum separation of clusters in SSC, and $\sim2$ times the value obtained for HRSC. These results can be considered as lower limits that can be improved once CHANCES spectroscopy is available.   

\subsection{Alignment between optical filaments and X-ray clusters}

For the X-ray clusters connected to filaments, we further investigated the morphology of the X-ray emission in relation to the underlying distribution of galaxies observed in the optical. To this end, we compared the inclination angle of the major axis of the X-ray emission with the projected orientation of the connected segment of the optical filaments.

To quantify the alignment between the X-ray morphology of the clusters and their associated filaments, we defined a normalized vector $\hat{v}_{x}$ that points in the direction of the inclination angle $\phi_x$ of the major axis of the X-ray cluster, and the normalized vector $\hat{v}_{fil}$, which points in the direction of the projected inclination angle $\phi_{fil}$ of the filament connected to the corresponding X-ray cluster. Following the definition used in \cite{altay2006influence}, the  measure of alignment between an X-ray cluster and a connected filament is

\begin{equation}
    |\cos(\phi_{x,fil})|=|\hat{v}_{x}\cdot\hat{v}_{fil}| \mbox{,}
\end{equation}

where $\phi_{x,fil}$ is the angle between $\hat{v}_{x}$ and $\hat{v}_{fil}$.

A $|\cos(\phi_{x,fil})|$ value close to $1$ implies a good filament-cluster alignment ($\phi_{x,fil}=0\text{ deg}$), while a value close to $0$ implies the contrary ($\phi_{x,fil}=90\text{ deg}$). We have found a strong trend that filaments connected to X-ray clusters preferentially align with the major axis of this cluster, as shown in the alignment distribution of Fig.~\ref{fig:fraction_alignments}, where at a distance of 1$r_{200}$ (0.8$r_{200}$-1.2$r_{200}$; dark blue line) $\sim 40 \%$ of the connected filaments have a $|\cos(\phi_{x,fil})| \sim 1$. If we increase the $|\cos(\phi_{x,fil})|$ range to $0.8$ this percentage reaches $\sim 55\%$. The remaining possible values follow an almost uniform distribution, with no evidence of another preferred direction. Note that we have calculated the distribution using four equally sized bins, each spanning an angular difference of 20.5$\ \text{deg}$, based on the angular covariance value $\sigma_{\alpha}$ used to determine the projected inclination angle of the detected filaments. In this way, our comparison is performed within the minimum angular resolution achievable. 

In order to quantify the detected alignment signal between clusters and filaments, we have compared the observed distribution with a uniform distribution of $|\cos(\phi_{x,fil})|$ (gray-dashed horizontal line in Fig.~\ref{fig:fraction_alignments}). The comparison was performed using the Chi-squared test with a 95\% significance level. We found that at a distance of 1$r_{200}$, the observed distribution yields a $\chi^{2}$ statistic of approximately 10.9, with a p-value of 0.012, indicating that the detected signal significantly deviates from a uniform distribution. As the distance from the cluster center increases, the distribution becomes increasingly similar to a uniform distribution ($\chi^{2}$ statistic of 0.8 and p-value of 0.85 at 1.6$r_{200}$, and $\chi^{2}$ statistic of 0.9 and p-value of 0.81 at 3.2$r_{200}$, see Fig.~\ref{fig:fraction_alignments}).

\begin{figure}
    \centering
    \includegraphics[width=1.\linewidth]{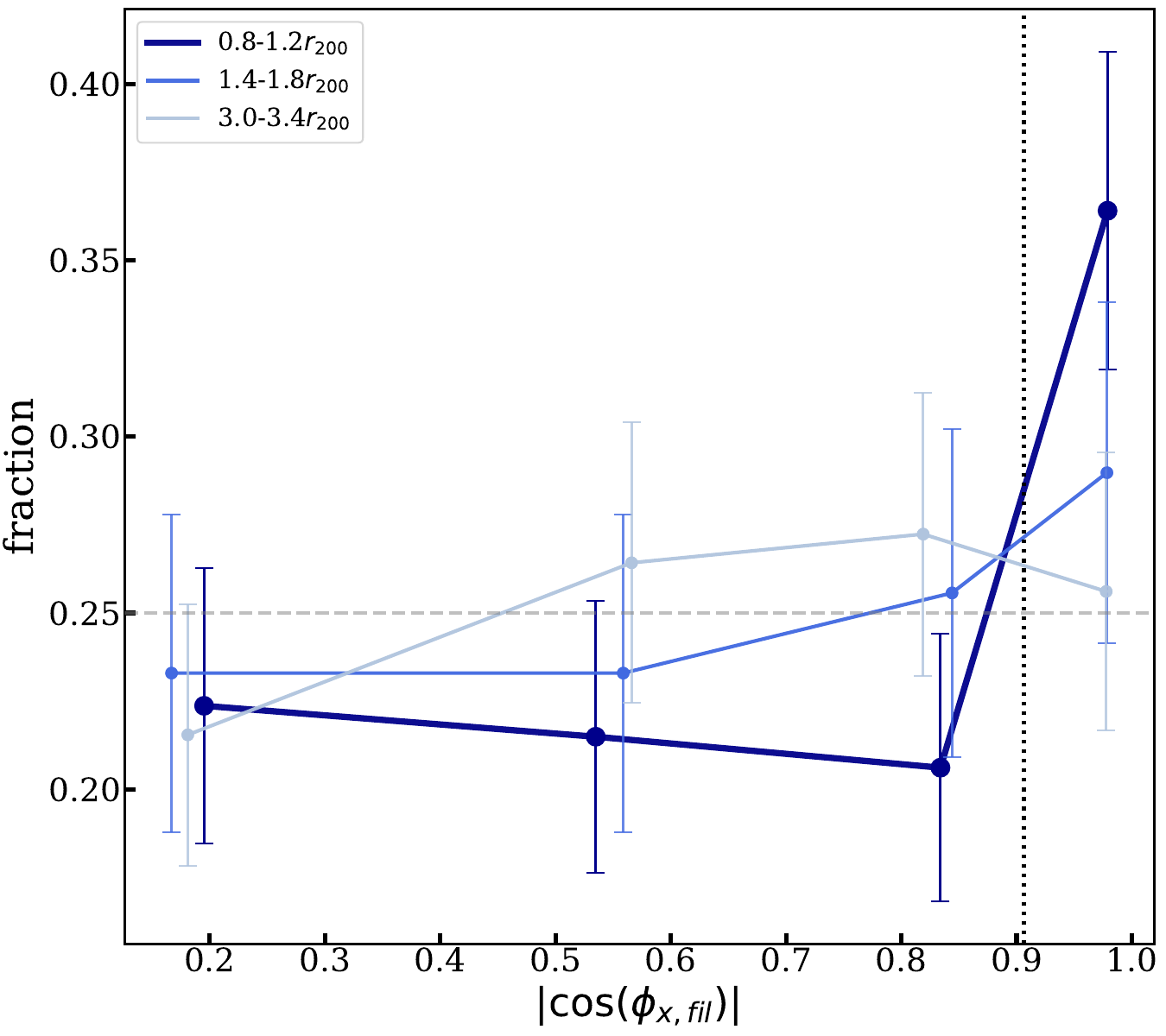}
     \caption{Distribution of the alignment between optical filaments connected to galaxy clusters and the major axis of each cluster as inferred from its X-ray emission. Fractions and 68\% confidence intervals were computed in equal-sized angular bins, following the method described in \cite{damsted2023codex}. Different colors indicate different projected cluster–filament distances used to compute the projected filament inclination. The gray-dashed horizontal line represents the uniform distribution, while the black-dotted vertical line indicates an alignment threshold of less than 25 degrees.}
    \label{fig:fraction_alignments}
\end{figure}

\section{Discussion and conclusions}\label{conclusions}

Mass accretion rates in galaxy clusters play a crucial role in galaxy evolution, including pre-processing effects \citep[e.g.,][]{sampaio2021}. A preferential accretion direction leads to an alignment between filaments and cluster shapes. Here, we present the first observational test of this correlation, showing that the alignment between filament elongation and cluster morphology provides a simplified statistical approach to studying the complex process of galaxy accretion.

Our results confirm a significant alignment between filaments and the major axes of galaxy clusters, {comparable} with findings from simulations. For instance, \cite{Kuchner2020} found that filaments tend to align with the major axis of the clusters in the inner region, a result that agrees with previous theoretical studies \citep{hahn2007evolution,youcai2009spin,ganeshaiah2018cosmic} and is further supported by the high fraction of alignment revealed in our analysis at ${1r_{200}}$ from the cluster center. Regardless of the latter, our analysis focuses on clusters within superclusters, using X-ray emission to trace their gravitational potential and photometric members to identify filaments, while simulations typically study isolated clusters in three dimensions based on dark matter and gas particle distribution. To enable a more direct comparison between our findings and simulations, studies employing X-ray mocks of superclusters are still required. We will conduct a similar analysis based on supercluster simulations, where X-ray mocks are being produced.

Despite clusters in the SSC and HRSC being closer to each other than clusters in the field, the influence of the supercluster environment does not appear to
significantly affect the cluster-filament alignment. We note the use of the X-ray emission as a tracer of the cluster shape and the lack of access to the
three-dimensional distribution of clusters and filaments. A control sample of isolated clusters is needed to understand how this correlation behaves in less dense environments. To address this, we will further test it by extending our analysis to the full CHANCES low-z cluster sample in future work. The spectroscopic follow-up by CHANCES will also provide improved mass accretion rate estimates, allowing for a more comprehensive analysis of the impact of cluster growth on galaxy evolution.

While our current data relies heavily on photometric redshifts to reconstruct the filament network in two dimensions, this does not reduce the reliability of the alignment statistics, as alignments with two-dimensional X-ray maps only work in the observer plane anyway, and it is anticipated that only the strongest filaments would dominate the alignment. With a more complete study, offered by the spectroscopy, we can fold in the connectivity to the picture.

Photometric redshifts have proven to be effective in tracing the most prominent components of the large-scale structure, as validated by comparisons with available spectroscopic data. The observed alignment of filaments detected in optical data with the X-ray elongation of clusters supports the idea that most of the mass accretion onto clusters occurs through these filaments. 

\begin{acknowledgements}

{We thank the anonymous referee for their careful and constructive review, which helped improve the clarity and quality of the manuscript.} 
{RBS} acknowledges support from the  Agencia Nacional de Investigaci\'on y Desarrollo (ANID)/Subdirección de Capital Humano/Doctorado Nacional/2023-21231017, Millennium 
Science Initiative Program NCN2024\_112,  Programa de Iniciación a la Investigación Científica
N° 029/2025 de la Universidad Técnica Federico Santa María, and ANID BASAL project FB210003.

YLJ acknowledges support from the Agencia Nacional de Investigaci\'on y Desarrollo (ANID) through Basal project FB210003, FONDECYT Regular projects 1241426 and 123044, Millennium Science Initiative Program NCN2024\_112, and Programa de Iniciación a la Investigación Científica N° 029/2025 de la Universidad Técnica Federico Santa María.

AF thanks FINCA 2609000, USM, UDA for the travel support. 

CPH acknowledges support from ANID through Fondecyt Regular project number 1252233 and ANID – MILENIO – NCN2024\_112.

HMH acknowledges support from the Agencia Nacional de Investigaci\'on y Desarrollo (ANID) through Fondecyt
project 3230176, ANID – MILENIO–NCN2024-112, and ANID BASAL project FB210003. 

AM acknowledges support from the ANID FONDECYT Regular
grant 1251882, from the ANID BASAL project FB210003, and funding
from the HORIZON-MSCA-2021-SE-01 Research and Innovation Programme
under the Marie Sklodowska-Curie grant agreement number 101086388.

UK acknowledges financial support from the UK Science and Technology Facilities Council (STFC; grant ref ST/T000171/1).

RS acknowledges financial support from FONDECYT Regular projects 1230441 and 1241426, and also gratefully acknowledges financial support from ANID – MILENIO – NCN2024\_112.

MAF acknowledges support from the Emergia program (EMERGIA20\_38888)
from Junta de Andalucía and University of Granada.

RD gratefully acknowledges support by the ANID BASAL project FB210003.

CL-D acknowledges a grant from the ESO Comité Mixto ORP037/2022, and the support from the Agencia Nacional de Investigación y Desarrollo (ANID) through Fondecyt project 3250511.

LSJ acknowledges the support from CNPq (308994/2021-3) and FAPESP (2011/51680-6).

F.A.-F. acknowledges support from FAPESP grants 2024/00822-5 and 2024/22842-8.

F.R.H. acknowledges support from FAPESP grants 2018/21661-9 and 2021/11345-5, and the S-PLUS Consortium for continuing to fund the T80-South operations and keeping the lights on.

E.I. gratefully acknowledge financial support from ANID - MILENIO - NCN2024\_112 and ANID FONDECYT Regular 1221846.

S.L. acknowledges support by FONDECYT grant 1231187.

Part of this work was supported by the Polish Ministry of Science and
Higher Education (MNiSW) grant DIR/WK/2018/12.
Part of this work was supported by grant 537 (pl0201-01) at the Poznań
Supercomputing and Networking Center.

GD acknowledges support by UKRI-STFC grants: ST/T003081/1 and ST/X001857/1. 

VMS acknowledge the support from ESO through the grants ORP026/2021, and the financial support from ANID - MILENIO - NCN2024\_112.

This work is partially based on data from eROSITA, the soft X-ray instrument aboard SRG, a joint Russian-German science mission supported by the Russian Space Agency (Roskosmos), in the interests of the Russian Academy of Sciences represented by its Space Research Institute (IKI), and the Deutsches Zentrum für Luft- und Raumfahrt (DLR). The SRG spacecraft was built by Lavochkin Association (NPOL) and its subcontractors and is operated by NPOL with support from the Max Planck Institute for Extraterrestrial Physics (MPE). The development and construction of the eROSITA X-ray instrument was led by MPE, with contributions from the Dr. Karl Remeis Observatory Bamberg \& ECAP (FAU Erlangen-Nuernberg), the University of Hamburg Observatory, the Leibniz Institute for Astrophysics Potsdam (AIP), and the Institute for Astronomy and Astrophysics of the University of Tübingen, with the support of DLR and the Max Planck Society.

\end{acknowledgements}

\bibliographystyle{aa} 
\bibliography{aa56957-25.bib}

\begin{appendix}\label{sec:appendix}

\onecolumn

\section{X-ray cluster properties}

Tables~\ref{tab:x_ray_table} and~\ref{tab:x_ray_table_HRSC} list the main properties of the galaxy clusters detected by eROSITA in the 0.6–2.3 keV band and used in this work.

\begin{table*}[h!]

\centering 
\caption{Properties of the 19 X-ray clusters in the SSC.}
\label{tab:x_ray_table}
\begin{threeparttable}

\begin{tabular}{ccccccccc}
\toprule
\hline \\
{(1)} & {(2)} &{ (3) }&{ (4)} & {(5)} & {(6)} & {(7)} & {(8)} & {(9)}
\\

RA (J2000) & Dec (J2000) & $z$ & $r_{200}$ & $M_{200}$  & $L_{x} $ & $a$ & $b$ & $\phi_{x}$ \\ 

(deg) & (deg) & & (Mpc) &  ($10^{14} \mbox{M}_{\odot}$) & ($10^{44}$ erg s$^{-1}$) & (arcmin) & (arcmin) & (deg) \\ \\ \hline \\

200.866 & -31.722 & 0.048 & 1.154 & 1.685 & 0.288 & 23.077 & 13.485 & 99.297 \\
194.167 & -30.357 & 0.055 & 1.638 & 4.904 & 1.542 & 23.071 & 10.082 & 90.257 \\
202.038 & -31.482 & 0.048 & 1.895 & 7.530 & 2.987 & 20.378 & 10.598 & 114.657 \\
194.537 & -28.514 & 0.066 & 1.053 & 1.301 & 0.196 & 19.932 & 9.228 & 110.027 \\
193.639 & -29.149 & 0.054 & 1.722 & 5.468 & 1.824 & 19.823 & 11.397 & 162.367 \\
193.102 & -31.267 & 0.054 & 1.529 & 3.899 & 1.076 & 15.827 & 9.373 & 98.757 \\
203.439 & -31.667 & 0.049 & 1.600 & 4.342 & 1.264 & 15.789 & 12.028 & 87.947 \\
195.959 & -31.410 & 0.055 & 0.895 & 0.770 & 0.085 & 15.678 & 10.128 & 42.037 \\
196.513 & -34.399 & 0.063 & 1.042 & 1.251 & 0.184 & 14.735 & 10.393 & 43.477 \\
203.108 & -33.074 & 0.049 & 1.400 & 3.003 & 0.711 & 12.645 & 11.380 & 170.867 \\
194.259 & -31.345 & 0.056 & 1.106 & 1.459 & 0.232 & 12.240 & 7.531 & 57.987 \\
199.884 & -33.480 & 0.046 & 1.066 & 1.298 & 0.191 & 11.428 & 7.626 & 59.357 \\
202.439 & -29.538 & 0.049 & 0.946 & 0.943 & 0.116 & 10.168 & 6.812 & 48.587 \\
203.655 & -35.271 & 0.052 & 0.796 & 0.541 & 0.049 & 9.386 & 6.156 & 47.727 \\
202.022 & -34.065 & 0.049 & 0.937 & 0.907 & 0.110 & 8.209 & 7.028 & 3.427 \\
202.374 & -28.311 & 0.034 & 0.639 & 0.279 & 0.017 & 12.788 & 9.123 & 92.087 \\
198.556 & -33.824 & 0.050 & 0.909 & 0.820 & 0.094 & 8.507 & 6.139 & 166.757 \\
200.405 & -35.823 & 0.050 & 0.866 & 0.706 & 0.074 & 8.491 & 7.671 & 63.977 \\
201.756 & -27.168 & 0.036 & 1.375 & 2.794 & 0.625 & 18.500 & 14.155 & 29.658 \\

\\
\hline 
\end{tabular}

\begin{tablenotes}[flushleft]
\small 
\item  {\textbf{Notes.} Columns are: (1) and (2): right ascension and declination of the cluster center, defined by the X-ray peak from the eROSITA detections (purple circles in Fig. \ref{fig:Shapley}); (3): redshift of the cluster; (4) and (5): radius and corresponding mass enclosing a mean density of 200 times the critical mass density of the Universe at the cluster redshift; (6): X-ray luminosity of the cluster; (7) and (8): major and minor axes of the ellipse fitted to the X-ray emission of the cluster; (9): inclination angle of the major axis (see Sect. \ref{subsec:x-ray_shape}).}

\end{tablenotes}
\end{threeparttable}

\end{table*}

\begin{table*}[h!]
\centering 

\caption{Same as {Table~\ref{tab:x_ray_table}} but for the clusters identified in the HRSC.}
\label{tab:x_ray_table_HRSC}
\begin{threeparttable}

\begin{tabular}{ccccccccc}
\toprule
\hline \\ 
{(1)} & {(2)} &{ (3) }&{ (4)} & {(5)} & {(6)} & {(7)} & {(8)} & {(9)}
\\
RA (J2000) & Dec (J2000) & $z$ & $r_{200}$ & $M_{200}$  & $L_{x} $ & $a$ & $b$ & $\phi_{x}$ \\ 

 (deg) & (deg) & & (Mpc) &  ($10^{14} \mbox{M}_{\odot}$) & ($10^{44}$ erg s$^{-1}$) & (arcmin) & (arcmin) & (deg) \\ \\ \hline \\

65.623 & -53.279 & 0.043 & 0.864 & 0.698 & 0.072 & 22.197 & 6.926 & 39.044 \\
53.629 & -53.618 & 0.065 & 1.014 & 1.155 & 0.163 & 18.062 & 11.314 & 124.794 \\
52.402 & -45.997 & 0.069 & 1.327 & 2.596 & 0.580 & 17.659 & 9.299 & 83.586 \\
59.071 & -53.856 & 0.038 & 0.745 & 0.446 & 0.036 & 17.606 & 7.950 & 101.114 \\
50.653 & -53.184 & 0.077 & 1.282 & 2.360 & 0.505 & 17.431 & 7.174 & 75.184 \\
50.415 & -45.483 & 0.072 & 1.247 & 2.162 & 0.437 & 17.117 & 9.452 & 129.886 \\
56.394 & -41.214 & 0.060 & 1.337 & 2.634 & 0.588 & 16.653 & 9.010 & 93.116 \\
49.744 & -53.906 & 0.054 & 0.942 & 0.915 & 0.112 & 16.604 & 7.522 & 20.254 \\
48.699 & -58.057 & 0.064 & 0.968 & 1.001 & 0.130 & 16.311 & 9.410 & 100.613 \\
56.459 & -56.976 & 0.059 & 1.252 & 2.159 & 0.430 & 15.960 & 11.344 & 169.563 \\
47.675 & -47.354 & 0.082 & 1.171 & 1.809 & 0.336 & 15.475 & 6.498 & 141.945 \\
55.133 & -55.079 & 0.044 & 1.027 & 1.173 & 0.163 & 14.379 & 11.010 & 17.844 \\
58.065 & -54.861 & 0.046 & 1.066 & 1.314 & 0.195 & 13.985 & 9.499 & 73.544 \\
52.605 & -52.547 & 0.060 & 1.466 & 3.464 & 0.900 & 13.969 & 11.746 & 8.205 \\
55.662 & -53.628 & 0.059 & 1.840 & 6.851 & 2.611 & 13.816 & 10.886 & 85.154 \\
55.385 & -41.024 & 0.062 & 0.856 & 0.692 & 0.073 & 13.439 & 6.583 & 92.726 \\
51.180 & -58.624 & 0.077 & 0.924 & 0.884 & 0.109 & 12.946 & 5.639 & 44.862 \\
62.348 & -59.587 & 0.058 & 0.952 & 0.948 & 0.119 & 12.880 & 6.695 & 84.332 \\
49.484 & -45.756 & 0.077 & 1.228 & 2.074 & 0.412 & 12.617 & 8.572 & 153.116 \\
57.082 & -45.508 & 0.070 & 1.080 & 1.402 & 0.222 & 12.507 & 7.169 & 38.246 \\
48.272 & -47.400 & 0.082 & 0.980 & 1.060 & 0.146 & 12.381 & 7.818 & 75.335 \\
53.490 & -39.061 & 0.063 & 1.006 & 1.123 & 0.155 & 11.765 & 9.191 & 158.716 \\
53.935 & -45.124 & 0.069 & 0.942 & 0.927 & 0.116 & 10.938 & 7.263 & 174.526 \\
48.595 & -45.395 & 0.077 & 1.571 & 4.344 & 1.310 & 10.925 & 9.996 & 72.356 \\
60.167 & -53.680 & 0.072 & 0.942 & 0.931 & 0.117 & 10.913 & 7.392 & 74.114 \\
55.148 & -45.705 & 0.070 & 1.194 & 1.893 & 0.355 & 10.627 & 7.472 & 88.186 \\
65.029 & -51.430 & 0.070 & 1.100 & 1.479 & 0.241 & 10.478 & 6.078 & 31.255 \\
50.583 & -41.338 & 0.063 & 1.366 & 2.814 & 0.653 & 10.305 & 10.152 & 119.746 \\
48.406 & -38.305 & 0.081 & 1.208 & 1.983 & 0.386 & 10.011 & 7.417 & 130.096 \\
50.642 & -49.254 & 0.070 & 0.887 & 0.777 & 0.088 & 9.819 & 4.657 & 76.715 \\
50.521 & -51.322 & 0.070 & 1.122 & 1.569 & 0.265 & 9.713 & 6.749 & 146.965 \\
65.700 & -51.512 & 0.043 & 1.086 & 1.385 & 0.210 & 9.301 & 7.254 & 157.285 \\
53.931 & -38.689 & 0.062 & 0.865 & 0.715 & 0.077 & 9.059 & 5.949 & 107.506 \\
49.095 & -50.957 & 0.074 & 0.934 & 0.909 & 0.113 & 8.988 & 7.618 & 72.395 \\
58.482 & -38.555 & 0.079 & 0.946 & 0.948 & 0.122 & 8.939 & 6.166 & 115.646 \\
59.766 & -59.597 & 0.062 & 0.806 & 0.577 & 0.055 & 8.723 & 6.508 & 113.682 \\
52.036 & -57.725 & 0.067 & 0.986 & 1.062 & 0.143 & 8.604 & 7.479 & 114.043 \\
51.313 & -56.591 & 0.060 & 1.953 & 8.204 & 3.467 & 8.489 & 6.541 & 124.393 \\
55.136 & -57.040 & 0.057 & 0.848 & 0.670 & 0.069 & 7.817 & 5.086 & 137.213 \\
50.264 & -56.213 & 0.083 & 1.040 & 1.266 & 0.192 & 7.596 & 5.600 & 102.503 \\
53.720 & -39.484 & 0.073 & 0.874 & 0.745 & 0.083 & 7.389 & 5.648 & 75.866 \\
55.700 & -46.298 & 0.070 & 1.052 & 1.296 & 0.196 & 7.329 & 6.339 & 40.266 \\
56.812 & -54.074 & 0.070 & 0.726 & 0.426 & 0.034 & 7.027 & 5.241 & 159.274 \\

\\
\hline 

\end{tabular}

\begin{tablenotes}[flushleft]
\small 
\item {\textbf{Notes.} Columns are as in Table \ref{tab:x_ray_table}.}

\end{tablenotes}
\end{threeparttable}
\end{table*}

\end{appendix}

\end{document}